# An Integrated EMT Small-Signal Stability Analysis Tool for Power Systems with High Inverter-Based Resource Penetration

Zihao Qin
*Purdue University*
West Lafayette, IN, USA
qin166@purdue.edu

Xiaonan Lu
*Purdue University*
West Lafayette, IN, USA
lu998@purdue.edu

Shuan Dong
*National Laboratory of the Rockies*
Golden, CO, USA
shuan.dong@nlr.gov

Jin Tan
*National Laboratory of the Rockies*
Golden, CO, USA
jin.tan@nlr.gov

***Abstract*—The ongoing replacement of synchronous generation by inverter-based resources (IBRs) introduces fast converter control dynamics whose characteristic frequencies extend beyond the classical electromechanical band. Conventional small-signal stability tools are commonly formulated in the phasor domain, representing electrical quantities as slowly varying phasors at the fundamental frequency, and therefore cannot resolve the sub-synchronous and converter-driven dynamics that increasingly arise in operation. Electromagnetic Transient (EMT) modeling captures these dynamics, but established EMT simulators produce time-domain waveforms rather than the modal indicators (eigenvalues, damping ratios, and participation factors) needed to assess stability risk. This paper presents EMT-SSA, an integrated tool suite for EMT-level small-signal analysis of IBR-rich power systems. From a standard PSS®E system snapshot (.raw/.dyr), it converts the model into an EMT representation with enhanced fidelity, solves an extended power flow for the steady-state equilibrium points of both the network and the device controllers, and linearizes a full-order EMT model to form the system state matrix, from which it produces eigenvalues, oscillation modes, and participation factor analysis results over a device library spanning synchronous generators, grid-following and grid-forming inverters, transmission lines, and loads. The EMT-SSA tool suite identifies poorly damped or unstable modes and attributes them to specific devices. The tool is demonstrated on the Kundur two-area system, with its modal results benchmarked against PSS®E NEVA.**

## I. Introduction

Modern bulk power systems are transitioning from synchronous-generator (SG) dominated operation toward one in which a growing share of generation is interfaced through power electronic inverters. Unlike synchronous machines, inverter-based resources (IBRs) impose dynamics governed by their inner current loops, phase-locked loops (PLLs), and outer control loops. These control-driven dynamics act on substantially faster timescales than the electromechanical oscillations that have historically defined power system stability, coupling stability phenomena that classical analysis treated as separable [1]. Grid functions and analyses, however, are largely designed on phasor-domain models, which represent electrical quantities as fundamental frequency phasors and therefore overlook the faster dynamics that IBRs introduce. Electromagnetic transient (EMT)-based analysis is increasingly needed to identify and diagnose them.

Ensuring reliable operation requires not only capturing these dynamics but analyzing them, such as identifying oscillatory modes, quantifying their damping, and tracing each to its root cause. Modal (small-signal) analysis provides this through the system's eigenvalues and the localization of each mode to its responsible devices via participation factors. Yet the platforms on which grid dynamic and small-signal studies are performed, including PSS®E [2], PSLF [3], PowerWorld [4], and TSAT [5], among others, are phasor-based. Built on fundamental-frequency phasor representations, these models reliably capture only dynamics that are slow relative to the fundamental frequency and lose fidelity across the faster range excited by IBRs. The limitation is not hypothetical. IBR-rich power grids with series-compensated devices along the line or low-strength interconnections have experienced sub-synchronous oscillation events in this range [6], emerging unforeseen issues since the small-signal tools operators rely on cannot flag them in advance. Commercial platforms have responded by adding increasingly detailed generic IBR models, but these remain built atop an unchanged phasor framework; as inverters replace synchronous generation, the underlying representation, not merely the device library, loses fidelity. The gap is one of modeling frameworks with extended functions, not of missing component models.

Electromagnetic-transient modeling directly resolves the fast IBR dynamics the phasor framework ignores, and several EMT environments are widely used for this purpose, including the commercial tools PSCAD/EMTDC [7] and EMTP [8], and the open-source ParaEMT [9]. These share the EMT fidelity sought here but differ in analysis method. As time-domain solvers, they numerically simulate the system and return waveforms rather than a modal description, so recovering eigenvalues, damping ratios, and participation factors requires either signal-processing of the responses or a separate small-signal module appended to the simulator [10]. Small-signal formulations that instead linearize a full EMT-level model and analyze it directly yield modal information natively [11]; these, however, have generally been assembled by hand for a specific study system, which limits routine use and lacks sufficient scalability. What remains is an automated path from a standard system case to direct EMT-level small-signal analysis.

This paper presents EMT-SSA, an integrated tool suite for EMT-based small-signal stability and oscillation analysis of IBR-rich power systems. From a standard PSS®E snapshot (.raw and .dyr), EMT-SSA converts the phasor-domain models to EMT representations with higher fidelity and solves an extended power flow that establishes the steady-state operating points of both the network and the device controllers. Then the tool suite linearizes a full-order EMT model about that steady-state operating point to form the system small-signal state matrix. From this matrix it produces eigenvalue, oscillation, and participation factor analyses over a device library spanning SG, grid-following (GFL) and grid-forming (GFM) inverters, transmission lines, and loads, identifying poorly damped and unstable modes and attributing each of those to the devices that drive it. Whereas time-domain EMT simulators reproduce the same fast dynamics but return only waveforms, EMT-SSA yields the eigenvalues, damping ratios, and participation factors directly. The principal contribution is an automated pipeline that delivers direct, EMT-level modal analysis for IBR-rich systems described by standard case data, without manual model construction.

The remainder of this paper is organized as follows. Section II presents the theoretical formulation underlying EMT-SSA, from device modeling in a common reference frame through system linearization and modal analysis. Section III describes the tool architecture and its constituent stages: case-data import, model library, model assembly, and output production. Section IV reports case studies on the Kundur two-area system, benchmarked against PSS®E NEVA. Section V concludes and outlines directions for future work.

## II. EMT-Level Small-Signal Model Formulation

EMT-SSA forms a full-order EMT model of the complete system (every SG, inverter, transmission line, and load represented by its own dynamic states) and linearizes it about the operating point supplied by the extended power flow to obtain the system state matrix for modal analysis. The formulation follows the structure detailed in [11]; the full per-device state equations are given there and are summarized rather than reproduced here. The model is developed in four steps: device modeling in a common reference frame, network and load modeling, system assembly and linearization, and modal analysis.

### A. Device Modeling in a Common Reference Frame

Each device, a SG, GFL inverter, or GFM inverter, is described by a set of nonlinear differential equations in its own d–q reference frame,

$$\dot{x}_c = f_c(x_c, v_b, \omega_{com}), \quad i_c = g_c(x_c) \tag{1}$$

where $x_c$ is the device state vector, $v_b$ the terminal-bus voltage, $\omega_{com}$ the common-frame frequency, and $i_c$ the injected current. The SG is represented by a round-rotor machine model, whose rotor and flux dynamics follow the standard formulation [12], augmented by an IEEEG1 governor and a lead–lag exciter [11]. The GFL and GFM inverters comprise, respectively, phase-locked loop and power synchronization dynamics, outer- and inner-loop controllers, and LCL-filter states; their full equations are given in [11]. Since each device is defined in its own frame, all devices are referred to a single common D–Q frame through the rotation [13]

$$x_D = x_d\cos\delta - x_q\sin\delta, \quad x_Q = x_d\sin\delta + x_q\cos\delta \tag{2}$$

where $\delta$ is the angle between the device frame and the common frame. This referral allows the individual device models to be interconnected through the network.

### B. Network and Load Modeling

Transmission lines and loads are represented as dynamic elements in the same common frame [13]. Each line is a series R–L branch contributing its inductor-current states, and each load is a constant-impedance R–L element contributing its own dynamic states. Linearized about the operating point and aggregated over all elements, the line and load subsystems take the state-space form:

$$\Delta\dot{i}_{line} = A_{line}\Delta i_{line} + B_{line}\Delta v_b + B_{line\omega}\Delta\omega_{com} \tag{3}$$

$$\Delta\dot{i}_{load} = A_{load}\Delta i_{load} + B_{load}\Delta v_b + B_{load\omega}\Delta\omega_{com} \tag{4}$$

where $\Delta\dot{i}_{line}$ and $\Delta\dot{i}_{load}$, respectively, denotes the line and load current deviation. $A_{(.)}$ and $B_{(.)}$ are, respectively, the state and input matrix coefficients. $\Delta v_b$ and $\Delta\omega_{com}$ are, respectively, small-signal deviations of terminal voltage and common-frame frequency.

### C. System Assembly and Linearization

The operating point about which the system is linearized is obtained from the extended power flow [11], which augments the conventional power-flow solution with the operating dynamics of the device controllers so that the steady-state values of the controller states, not only the bus voltages and power injections, are determined. In this formulation the system frequency becomes an additional variable, and each bus is reclassified according to the control mode of its connected device, with grid-forming buses regulating voltage and frequency. Linearizing each device about this operating point gives a per-device small-signal model that, aggregated over all units, yields

$$\Delta\dot{x}_{gen} = A_{gen}\Delta x_{gen} + B_{gen}\Delta v_b + B_{gen\omega}\Delta\omega_{com} \tag{5a}$$

$$\Delta i_{gen} = C_{gen}\Delta x_{gen} \tag{5b}$$

The generator, line, and load subsystems are coupled through the network, whose terminal-voltage relation is

$$\Delta v_b = N_{gen}\Delta i_{gen} + N_{load}\Delta i_{load} + N_{line}\Delta i_{line} \tag{6}$$

where the matrices $N_{gen}$, $N_{load}$, and $N_{line}$ capture the interconnection, including the virtual resistances introduced for numerical conditioning. Substituting (6) into the device, line, and load models (1), (3)–(5) eliminates $\Delta v_b$ and combines all states into a single system,

$$\Delta\dot{x} = A_{sys}\Delta x \tag{7}$$

where $x$ aggregates the states of all generators, inverters, lines, and loads, and $A_{sys}$ is the full-order EMT small-signal state matrix on which the analysis operates.

*D. Modal Analysis*

The oscillatory behavior of the system is characterized by the eigenvalues of $A_{sys}$. For the i-th mode, with right and left eigenvectors $\varphi_i$ and $\psi_i$,

$$A_{sys}\phi_i = \lambda_i\phi_i, \quad \lambda_i = \sigma_i + j\omega_i \tag{8}$$

The oscillation frequency and damping ratio are

$$f_i = |\omega_i| / 2\pi, \quad \zeta_i = -\sigma_i / \sqrt{(\sigma_i^2 + \omega_i^2)} \tag{9}$$

A mode is stable when $\sigma_i < 0$; the system is small-signal stable when every mode satisfies this, and the least-damped modes are of primary interest. To identify the devices driving a given mode, the participation factor of the k-th state in the i-th mode is

$$p_{ki} = |\psi_{ik}||\phi_{ki}| / \Sigma_l|\psi_{il}||\phi_{li}| \tag{10}$$

which localizes each mode to its dominant states, and hence to the responsible devices [14]. Equations (8)–(10) produce the eigenvalue, oscillation, and participation-factor results reported by the tool (Section III-D).

## III. EMT-SSA Tool Architecture

EMT-SSA is organized as a modular framework of self-contained stages that pass data forward, so that new device models or analysis routines can be added without disturbing the rest of the pipeline. Four stages carry a case from standard input data to modal results, as illustrated in Fig. 1: case-data import, the model library, model assembly, and output production. This section describes each stage; the linearization and eigen-analysis that operate on the assembled model are formalized in Section II.

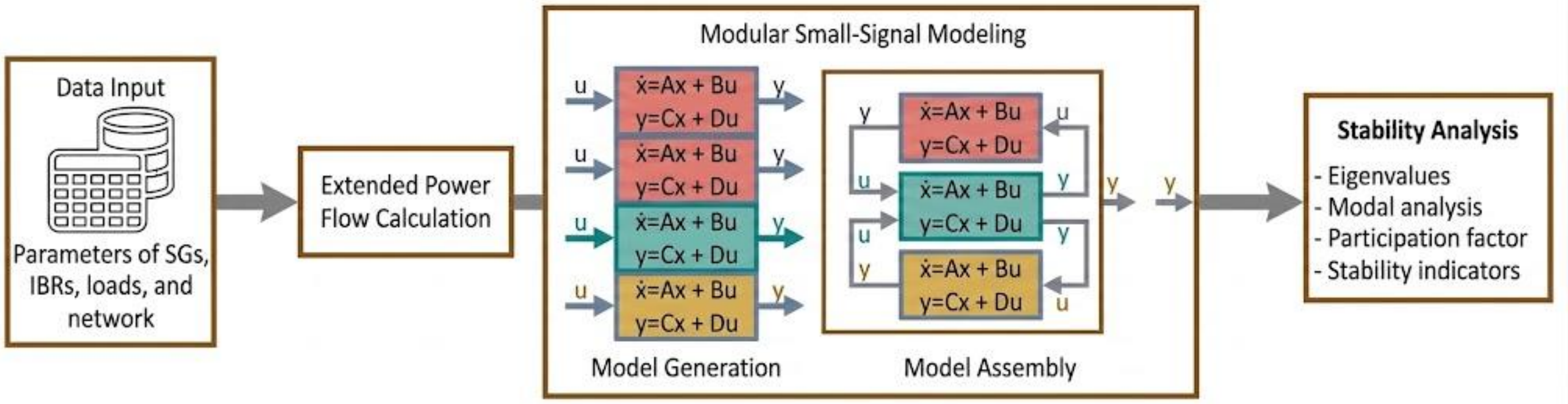

Fig. 1. EMT-SSA architecture: from a standard PSS®E case to eigenvalue, oscillation, and participation factor result.

*A. Case Data Import*

EMT-SSA takes standard PSS®E .raw and .dyr files as input and consolidates them into a single structured case that bundles the network and dynamic data in one record, giving one place to inspect or modify parameters. The .raw file supplies the bus, branch, transformer, load, and machine records that define the network and its operating configuration, while the .dyr file supplies the dynamic model assignments and parameters for each machine and controller. Each PSS®E dynamic model is mapped to its counterpart in the EMT-SSA device library. A preprocessing step then reduces the records to an assembly-ready form: machines on a common bus are merged into an equivalent unit, and parallel circuits are combined. Throughout, the original PSS®E bus indexing and per-unit conventions are preserved, so that every state and mode in the final results can be traced back to the named buses and devices of the source case for inspection.

*B. Model Library*

The model library is the collection of interchangeable component models from which any case is built. It currently provides five device models spanning the generation and interface technologies of IBR-rich systems, together with network and load elements:

- SG: a round-rotor machine with rotor angle and speed and transient/sub-transient flux states, augmented by an IEEEG1 governor and a SEXS exciter.
- GFL inverter, two variants: an inner-current-control model, and a model that adds an outer power-control loop. Both include the PLL, the inner d–q current PI loops, and the LCL-filter states.
- GFM inverter, two variants: droop-based and virtual-synchronous-machine (VSM). Both regulate frequency and voltage autonomously through inner voltage and current loops and include the LCL-filter states.
- Transmission lines: dynamic R–L branches, each contributing inductor-current states.
- Loads: constant-impedance elements, each contributing its own dynamic states.

The library is extensible: additional device models can be introduced without altering the surrounding stages.

### C. Model Assembly

In the assembly stage, each component named in the imported case is instantiated from the library and interconnected to form the full system model. Every inverter is placed on its own virtual bus through a short coupling branch, leaving the underlying transmission topology unchanged, and the transmission lines and loads are represented as the dynamic R–L elements defined in Section II, through which the components interconnect. The resulting interconnected model is nonlinear; its steady-state operating point is obtained from the extended power flow [11], which solves for the steady state of both the power network and the device controllers, so that the initial values of the controller states, not only bus voltages and injections, are consistent with the operating condition. Linearizing this model about the operating point yields the system state matrix on which the next stage operates; the linearization is developed in Section II.

### D. Output Production

From the assembled and linearized model, EMT-SSA produces three analyses. Eigenvalue analysis returns the full spectrum of the system state matrix and the corresponding stability assessment, taken from the rightmost eigenvalue. Oscillation analysis characterizes each oscillatory mode by its frequency and damping ratio and identifies the modes of interest. Participation-factor analysis localizes each mode to its dominant states, and thereby to the specific device or network element responsible, providing the root-cause information that distinguishes a modal result from a bare eigenvalue list.

## IV. Case Study

EMT-SSA is demonstrated on the Kundur test system. The Kundur two-area system, a canonical benchmark for inter-area electromechanical oscillation, is used to establish the correctness of the tool's output against well-understood behavior and against PSS®E NEVA. The case follows the workflow of the tool: the extended power flow establishes the operating point, after which eigenvalue, oscillation, and participation-factor analyses are produced.

### A. Kundur Two-Area System

The Kundur system comprises two areas, each with two SGs, joined by a double-circuit tie, with the system loads placed near the tie [14]; its single-line diagram is shown in Fig. 2. Each generator is represented by a round-rotor machine model with an IEEEG1 governor and an SEXS exciter, and the transmission lines and loads are modeled as dynamic elements as described in Section III. The case uses the standard parameter set [14], which exhibits a lightly damped inter-area mode and thus provides a concrete, well-documented target against which the EMT-SSA tool's output can be read.

#### 1) Power Flow Validation

Before any modal result is reported, the operating point produced by the extended power flow is verified against PSS®E [2]. Since EMT-SSA preserves the source case's bus indexing and per-unit conventions, the two solutions are compared bus by bus without case-specific tuning. Across all buses the voltage magnitudes agree to within $7\times10^{-5}$ p.u. and the bus angles to within 0.06°, as shown in Fig. 3. The extended power flow therefore reproduces the PSS®E operating point, establishing a consistent starting point for linearization.

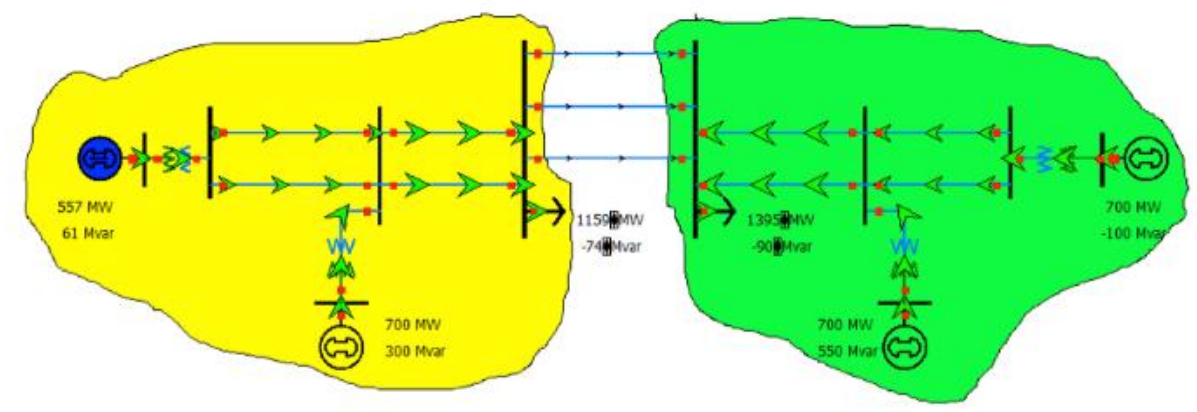

Fig. 2. Kundur two-area system [15].

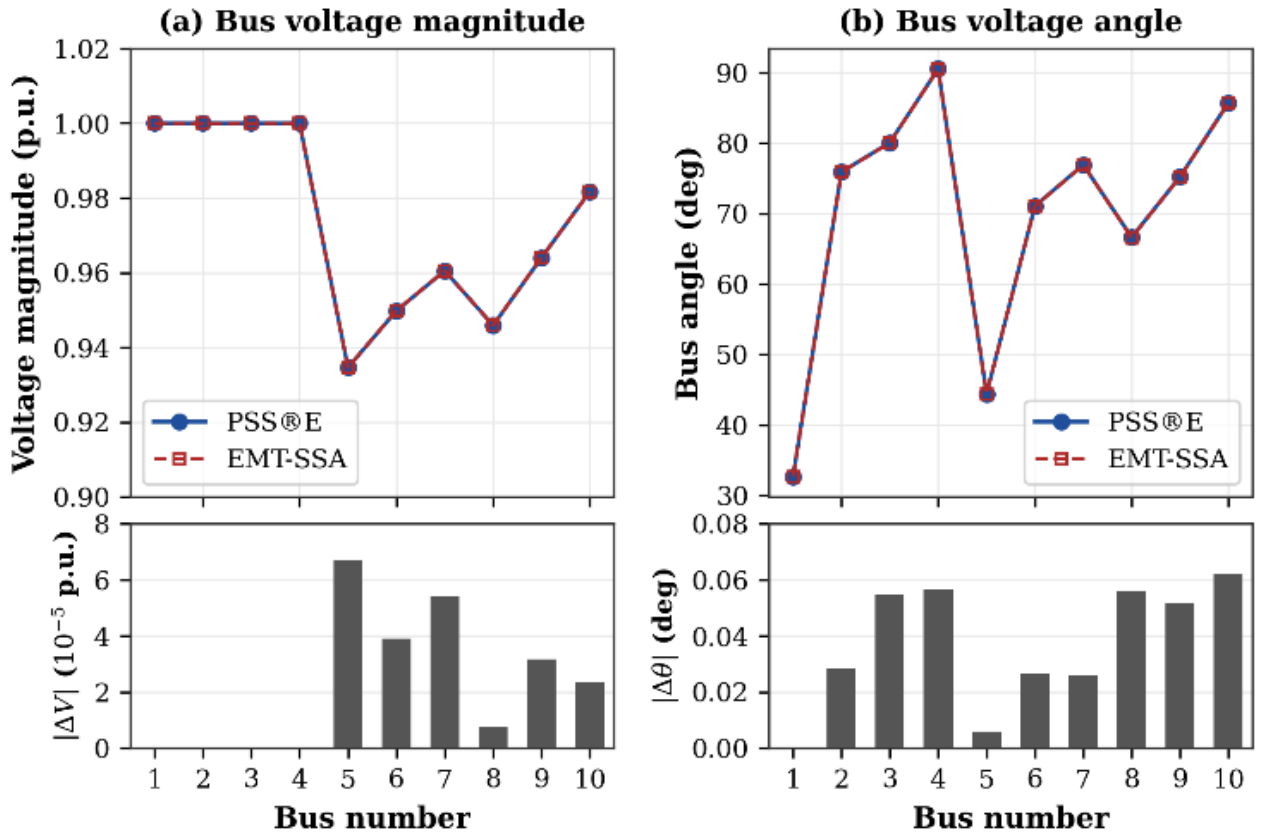

Fig. 3. EMT-SSA versus PSS®E power-flow solution for the Kundur system: (a) bus voltage magnitude and its difference; (b) bus voltage angle and its difference.

#### 2) Eigenvalue Analysis

Linearizing about this operating point yields the system state matrix, whose eigenvalues are shown in Fig. 4. The spectrum exhibits a clear two-timescale structure. One group of heavily damped modes lies near the fundamental frequency (imaginary part ≈ ±377 rad/s, i.e., ≈60 Hz) with large negative real parts; these are the high-frequency electromagnetic modes that the EMT-level model resolves. A second group lies near the imaginary axis at low frequency: the electromechanical modes. All modes have negative real parts: the system is small-signal stable, and the least-damped modes are the low-frequency electromechanical modes examined next.

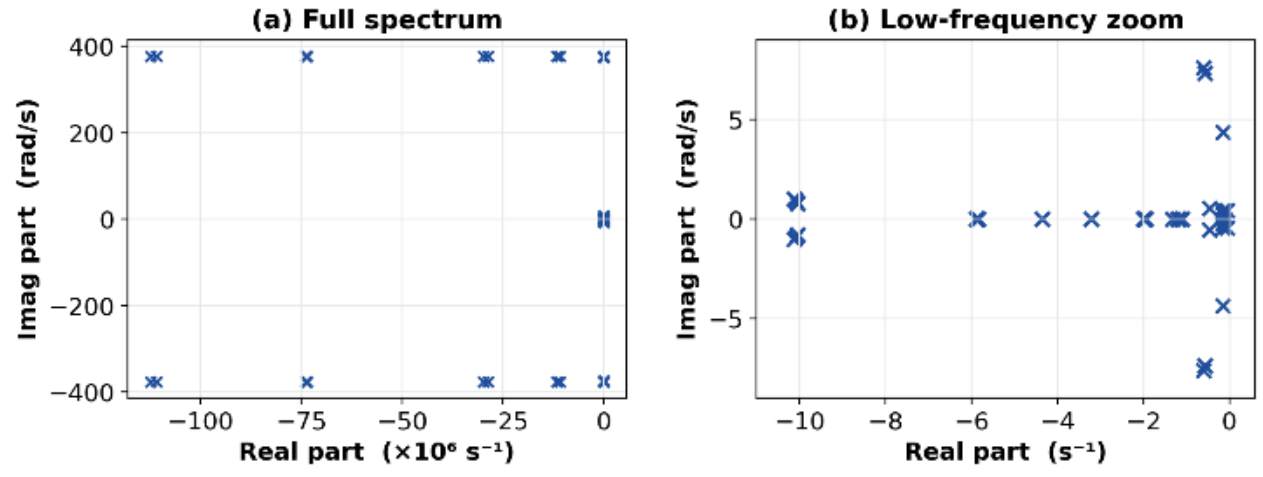

Fig. 4. Eigenvalue spectrum of the Kundur system: (a) full spectrum; (b) low-frequency zoom.

#### 3) Oscillation and Participation Factor Analysis

Table I lists representative modes spanning the spectrum, each characterized by its oscillation frequency and damping ratio and attributed to its dominant machines through participation factors. The tool reproduces the canonical Kundur behavior: a lightly damped inter-area mode near 0.70 Hz (ζ = 0.035) whose participation spans generators in both areas, together with local modes near 1.2 Hz. It additionally resolves the heavily damped ≈60 Hz electromagnetic modes that phasor-domain analysis does not represent. For every mode, the participation factors localize the oscillation to specific machines, for instance, the inter-area mode to SG@1, SG@3, and SG@4. This provides the root-cause information that distinguishes a modal result from a bare eigenvalue list.

TABLE I. SELECTED MODES OF THE KUNDUR SYSTEM

| Mode type | *f* (Hz) | ζ | Dominant machines |
|---|---|---|---|
| Electromagnetic (≈60 Hz) | 59.95 | 0.076 | Network |
| Machine (≈60 Hz) | 59.94 | 0.038 | SG@2 |
| Inter-area | 0.696 | 0.035 | SG@1, @3, @4 |
| Local | 1.170 | 0.078 | SG@1, @2 |
| Local | 1.211 | 0.079 | SG@3, @4 |
| Well-damped local | 0.160 | 0.995 | SG@1, @3, @4 |

*4) Comparison with PSS®E NEVA*

The lightly damped electromechanical modes are compared against PSS®E NEVA [2] on the same case (Table II). The inter-area and local modes appear in both tools with damping ratios of comparable magnitude, while the modal frequencies differ by up to about 11%. This offset originates in the load representation: EMT-SSA currently models loads as constant-impedance R–L elements, which at buses with net capacitive reactive load yield a negative equivalent inductance. That same representation produces the small number of spurious right-half-plane eigenvalues identified as numerical artifacts and excluded from the reported spectrum, and it is the principal modeling difference from NEVA's load treatment that shifts the electromechanical frequencies. Reconciling the load representation and adding a constant-power load model are planned (Section V). NEVA does not represent the ≈60 Hz electromagnetic modes, which EMT-SSA resolves directly.

TABLE II. *ELECTROMECHANICAL MODES: EMT-SSA VERSUS PSS®E NEVA*

| Mode | f (Hz) | | ζ | |
|---|---|---|---|---|
| | NEVA | EMT-SSA | NEVA | EMT-SSA |
| **Inter-area** | **0.627** | **0.696** | **0.038** | **0.035** |
| **Local** | **1.100** | **1.170** | **0.094** | **0.078** |
| **Local** | **1.133** | **1.211** | **0.098** | **0.079** |

## V. CONCLUSION AND FUTURE WORK

This paper presented EMT-SSA, an integrated tool suite for EMT-based small-signal stability and oscillation analysis of power systems with high penetration of IBRs. From a standard PSS®E case, the tool builds a full-order EMT model through automated data import, extended power flow, and model assembly, then linearizes it and analyzes the resulting state matrix to produce eigenvalue, oscillation, and participation-factor results. On the Kundur two-area benchmark, EMT-SSA reproduced the canonical electromechanical behavior, a lightly damped inter-area mode near 0.7 Hz and local modes near 1.2 Hz, and additionally resolved the heavily damped high-frequency electromagnetic modes that phasor-domain analysis does not represent, attributing every mode to its responsible machines through participation factors. Its electromechanical modes were benchmarked against PSS®E NEVA, showing damping ratios of comparable magnitude and frequencies differing by up to about 11%, a gap traced to the load representation. The extended power flow matched the PSS®E operating point to within $7\times10^{-5}$ p.u. in voltage magnitude and 0.06° in angle, confirming a consistent basis for linearization.

Future work will broaden the device library and validate it on IBR-rich case studies, and will scale the tool to larger interconnections through accelerated and reduced-order linearization and eigen-solution methods. Scalability to larger systems has already been verified; a detailed large-scale modal study, including the WECC 240-bus model [16], will be reported in future work.

## ACKNOWLEDGMENT

This work was authored in part by the National Lab oratory of the Rockies (NLR) for the U.S. Department of Energy (DOE), operated under Contract No. DE-AC36-08GO28308. Funding for the NLR authors provided by U.S. Department of Energy Office of Critical Minerals and Energy Innovation Integrated Energy Systems Office. The views expressed in the article do not necessarily represent the views of the DOE or the U.S. Government. The U.S. Government retains and the publisher, by accepting the article for publication, acknowledges that the U.S. Government retains a nonexclusive, paid-up, irrevocable, worldwide license to publish or reproduce the published form of this work, or allow others to do so, for U.S. Government purposes.